\documentclass[twocolumn]{webofc}
\usepackage[varg]{txfonts}   
\usepackage{xcolor} 
\begin{document}
\title{Power coupled to the slow wave resonance cone in cold plasma}
%
%

\author{\firstname{Wouter} \lastname{Tierens}\inst{1}\fnsep\thanks{\email{tierenswv@ornl.gov}}
}

\institute{Oak Ridge National Laboratory, 1 Bethel Valley Road, Oak Ridge, TN 37830, United States of America}

\abstract{%
We consider the power coupled to the plasma from a cylindrical source emitting resonance cones, propagative slow waves which exist in low density magnetized plasmas when the signs of the Stix parameters $S$ and $P$ differ. In this work, we calculate this power for the first time for any collisionality, and show that it remains finite in the cold collisionless limit, even as the radiofrequency electric field itself, and the associated power absorption density, is singular. We give the main parameter dependencies of the coupled power.
}
\maketitle
\section{Introduction}

A resonance cone is a near-electrostatic propagative slow wave which exists when the Stix parameter \cite{stix1992waves} $S$ is positive and the Stix parameter $P$ is negative. \textcolor{black}{Recent experiments suggest its presence even in current tokamaks} \cite{paulus2025icrf}, and it may play a more important role in larger future devices with lower edge densities. In previous work, we gave the exact analytic description of resonance cone emission in a cold plasma from a cylindrical source with radius $r_s$, both for the collisional \cite{tierens2024slow} and the collisionless cases \cite{tierens2023resonance}. We showed that the electric field is singular in the cold collisionless limit (just like what happens at the lower hybrid resonance).

\begin{figure*}
    \centering
    \includegraphics[width=0.49\linewidth]{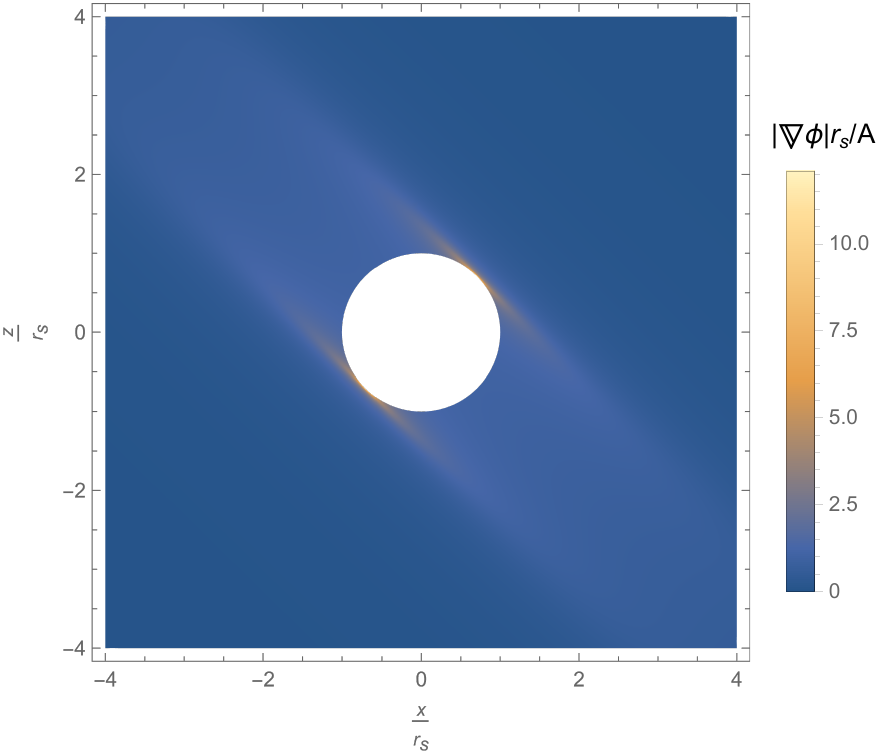}
    \includegraphics[width=0.49\linewidth]{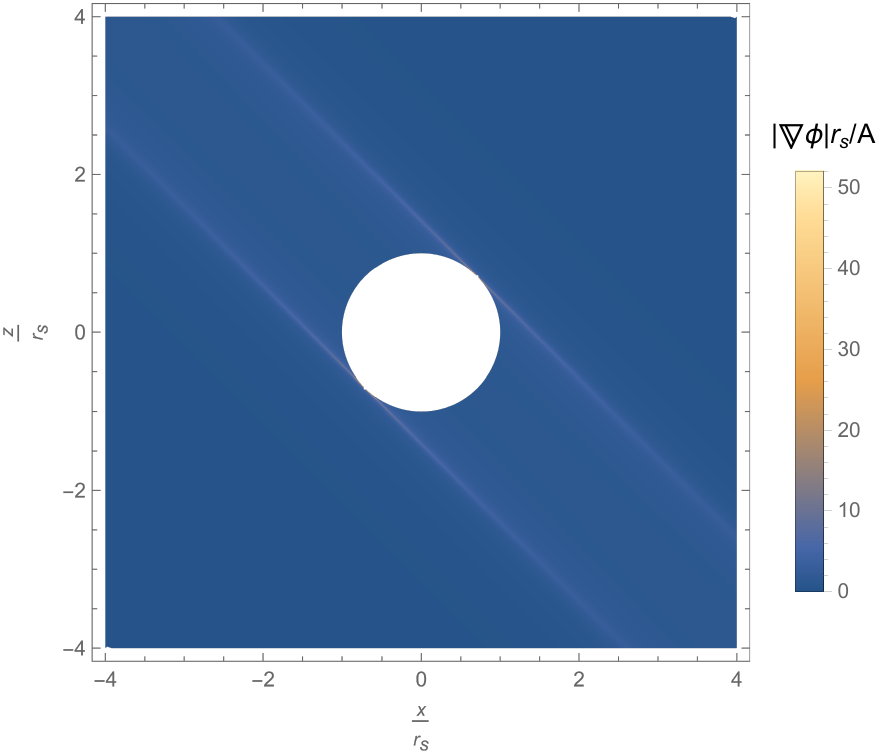}
    \caption{Left: electric field strength $|\nabla \phi|$ associated with the potential $\phi=A F_{-1}\left(\frac{x+s z}{r_s}\right)$ and $s=1+i/5$, normalized to $A/r_s$. The resonance cone, being preferentially emitted from the two tangent points, is clearly visible. Right: the same scenario at $s=1+i/50$, closer to the collisionless limit. The electric field is far more sharply peaked around the tangent points.}
    \label{fig:1}
\end{figure*}

\textcolor{black}{In this work, like in \cite{tierens2023resonance,tierens2024slow}, we consider the electrostatic description of the propagative slow wave in 2D, where $\boldsymbol{E}=-\nabla\phi$ and the potential $\phi$ obeys \cite{bellan2008fundamentals}
\begin{align}
    \frac{\partial^2}{\partial_z^2} \phi = \frac{-S}{P}\frac{\partial^2}{\partial_x^2} \phi \label{tdwe}
\end{align}
where $S,P$ are the usual Stix parameters \cite{stix1992waves}.}
We consider the collisional solution of (\ref{tdwe}) given in \cite{tierens2024slow}, where the potential at the source surface is $A \exp(-i m\theta)$ with $\theta$ the azimuthal angle and $A$ having units of potential [V], and $m>0$ is the (integer) mode number. This solution has the (complex) potential $\phi = A F_{-m}\left(\frac{x+s z}{r_s}\right)$ with
\begin{align}
    F_{-m}\left(\xi\right)= \left(\frac{\xi \left(\sqrt{1-\frac{s^2+1}{\xi^2}}-1\right)}{\sqrt{-s^2}-1}\right)^m \label{defPhi}
\end{align}
where $s=\sqrt{-S/P}$, the quantity which governs the resonance cone angle \cite{bellan2008fundamentals}. $S, P$ and $s$ are complex in the collisional case ($\Re(s)>0$, $\Im(s)>0$). The associated electric field strength $|\nabla \phi|$ is shown in figure \ref{fig:1} for $s=1+i/5$ and $s=1+i/50$, illustrating how the behaviour changes as the collisionless limit $\Im(s)\rightarrow 0^+$ is approached.

We now consider the total power coupled to the plasma due to this wave field. That power is \cite{brambilla1998kinetic}
\begin{align}
    \frac{\omega  \epsilon _0 |A|^2}{2 r_s^2} \left(|s|^2 \Im(P)+\Im(S)\right) \iint_{x^2+z^2>r_s^2} \left|F_{-m}'\left(\frac{x+s z}{r_s}\right)\right|^2 dx dz \label{firstIntegral}
\end{align}
We are interested in the behaviour of this integral for any $\Im(s)>0$, but especially in the $\Im(s)\rightarrow 0^+$ collisionless limit, where the field strength $| \nabla \phi|$ is singular.

\section{Evaluation of the power integral}
\begin{figure*}
    \centering
    \includegraphics[width=0.33\linewidth]{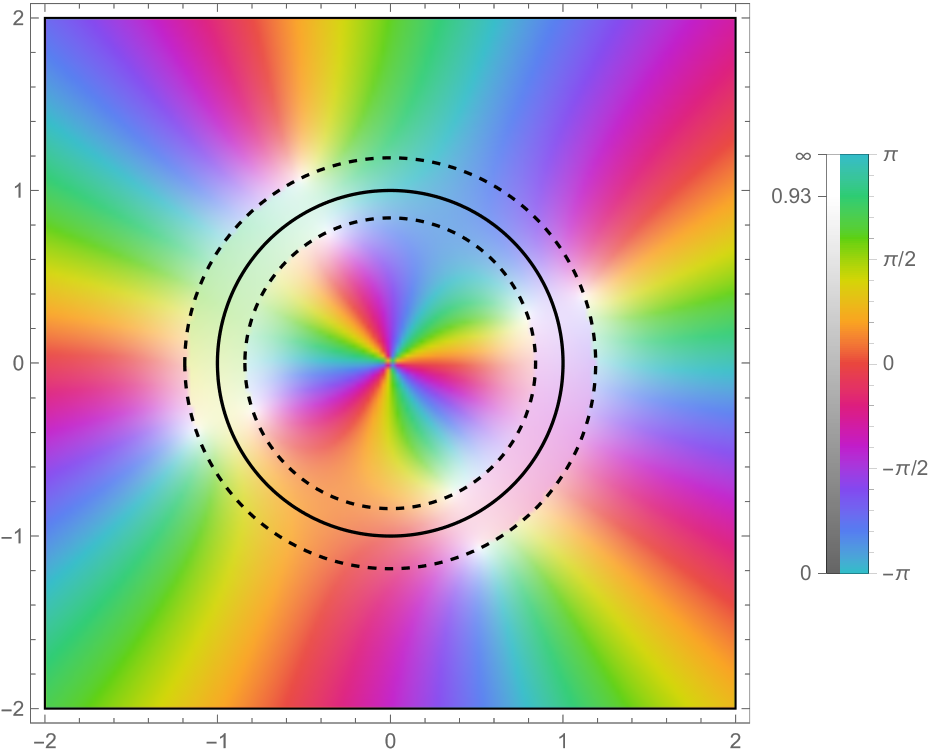}\includegraphics[width=0.33\linewidth]{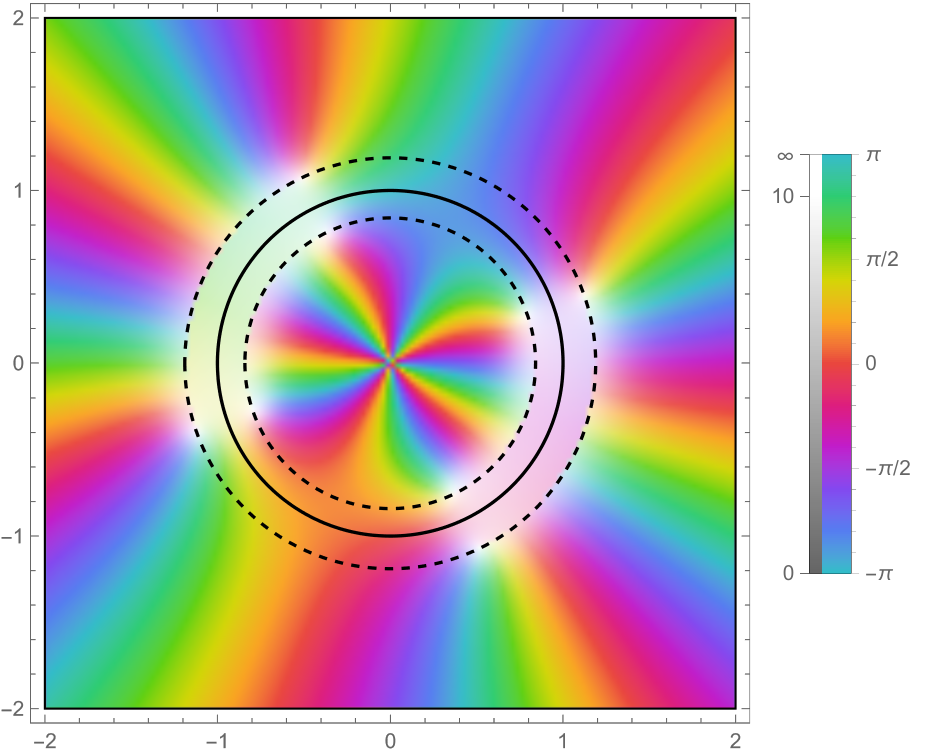}\includegraphics[width=0.33\linewidth]{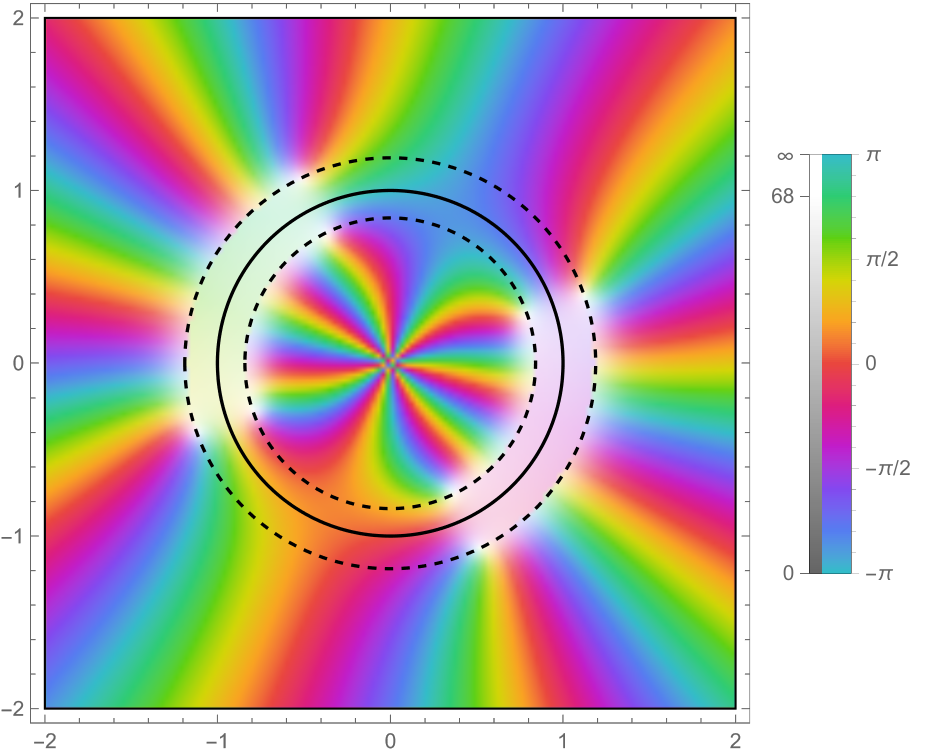}
    \caption{Complex plots of the integrand of (\ref{eq6}) for $m=1$ (left), $m=2$ (middle), and $m=3$ (right). The unit circle contour integration path is the solid black circle. The boundary of the annulus in which the integrand is holomorphic, i.e. $|\gamma|=r_a$ and $|\gamma|=R_a$, are shown as dashed circles.}
    \label{fig:m123}
\end{figure*}

Switching to polar coordinates and introducing the dimensionless radius $\rho=r/r_s$,
\begin{align}
    \frac{\omega \epsilon_0}{2}\frac{|A|^2}{r_s^2}(|s|^2 \Im(P)+\Im(S))\nonumber\\
    \cdot\int_{-\pi}^{\pi} \int_{r_s}^{\infty} r \left|F_{-m} '\left(\frac{r\cos(\theta)+s r \sin(\theta)}{r_s}\right)\right|^2 dr d\theta  \nonumber \\
    =
    \frac{\omega \epsilon_0}{2} |A|^2 (|s|^2 \Im(P)+\Im(S)) \nonumber\\ \cdot \int_{-\pi}^{\pi} \int_{1}^{\infty} \rho \left|F_{-m} '\left(\rho(\cos(\theta)+s \sin(\theta))\right)\right|^2 d\rho d\theta \label{ipol}
\end{align}
Let us start with the $\theta$ integration. First, we expand the absolute value squared
\begin{align}
      \left|F_{-m} '\left(\rho(\cos(\theta)+s \sin(\theta))\right)\right|^2  = \nonumber\\
    F_{-m} '\left(\rho(\cos(\theta)+s \sin(\theta))\right) G_{-m} '\left(\rho(\cos(\theta)+\overline{s} \sin(\theta))\right) \label{absExpand}
\end{align}
where
\begin{align}
    G_{-m}\left(\xi\right)= \left(\frac{\xi \left(\sqrt{1-\frac{\overline{s}^2+1}{\xi^2}}-1\right)}{\sqrt{-\overline{s}^2}-1}\right)^m \label{defGm}
\end{align}
Note the complex conjugate $\overline{s}$, and note that the identity (\ref{absExpand}) holds for real $\rho>1$ and real $\theta$. Allowing complex $\theta$, the rhs. of (\ref{absExpand}) is (at least locally) a holomorphic function of $\theta$, and the tools of complex analysis can be applied to integrate it. We convert the integral over $\theta$ to a contour integral around the unit circle using the standard trick
\begin{align}
    \int_{-\pi}^{\pi} f(\sin(\theta),\cos(\theta))d\theta = \oint\frac{1}{i\gamma} f\left(\frac{i}{2 \gamma }-\frac{i \gamma }{2},\frac{\gamma }{2}+\frac{1}{2 \gamma }\right) d\gamma \nonumber
\end{align}
which yields
\begin{align}
      \oint \frac{-i}{\gamma }F_{-m} '\left(\rho\left(\frac{1}{2} \left(\gamma +\frac{1}{\gamma }\right)+s \frac{i}{2} \left(\frac{1}{\gamma }-\gamma \right)\right)\right) \nonumber \\
      \cdot G_{-m} '\left(\rho\left(\frac{1}{2} \left(\gamma +\frac{1}{\gamma }\right)+\overline{s} \frac{i}{2} \left(\frac{1}{\gamma }-\gamma \right)\right)\right) d\gamma \label{eq6}
\end{align}
The first factor has poles that lie on a circle $|\gamma|=r_a$ with $r_a<1$. The second factor has poles that lie on a circle $|\gamma|=R_a$ with $R_a=1/r_a$. The integrand of (\ref{eq6}) is holomorphic in an annulus $r_a<|\gamma|<R_a$ surrounding the unit circle. Some examples are shown in figure \ref{fig:m123}.

We aim to compute the Laurent series on this annulus, which will let us evaluate the contour integral. First, recall that the Jacobi polynomials \cite{szeg1939orthogonal} $P_n^{(m,0)}(z)$ have the generating function
\begin{align}
    \sum_{n=0}^{\infty} t^n P_n^{(m,0)}(z) &= \frac{2^m \left(\sqrt{t^2-2 t z+1}-t+1\right)^{-m}}{\sqrt{t^2-2 t z+1}} \label{JacobiGF}
\end{align}
Two factors in the integrand of (\ref{eq6}) may be expanded in series of this form:
\begin{align}
    F_{-m} '\left(\rho\left(\frac{1}{2} \left(\gamma +\frac{1}{\gamma }\right)+s \frac{i}{2} \left(\frac{1}{\gamma }-\gamma \right)\right)\right) = \nonumber\\
    \frac{2 (-i)^{m+1} m }{(\sqrt{-s^2}-1)^m} \sum_{n=1}^{\infty} \frac{(s-i)^{n+m-1}}{(s+i)^n\gamma^{2n}} \rho ^{-m-1} P_{n-1}^{(m,0)}\left(1-\frac{2}{\rho ^2}\right)  \label{Fexpand}
\end{align}
which, being a Taylor series around $\gamma=\infty$, converges for $|\gamma|>r_a$, and
\begin{align}
    G_{-m} '\left(\rho\left(\frac{1}{2} \left(\gamma +\frac{1}{\gamma }\right)+\overline{s} \frac{i}{2} \left(\frac{1}{\gamma }-\gamma \right)\right)\right)= \nonumber\\  \frac{2 i^{m+1} m}{(\sqrt{-\overline{s}^2}-1)^m}  \sum_{n=1}^{\infty} \frac{(\overline{s}+i)^{n+m-1}}{(\overline{s}-i)^n} \gamma^{2n} \rho ^{-m-1} P_{n-1}^{(m,0)}\left(1-\frac{2}{\rho ^2}\right) \label{Gexpand}
\end{align}
which, being a Taylor series around $\gamma=0$, converges for $|\gamma|<R_a$. 
Note that the coefficient in $\gamma^{-2n}$ in (\ref{Fexpand}) is the complex conjugate of the coefficient in $\gamma^{2n}$ in (\ref{Gexpand}). It is thanks to this that our final result will be real and positive.

\textcolor{black}{Briefly expanding on the derivation of (\ref{Fexpand}) and (\ref{Gexpand}), we consider (\ref{JacobiGF}) with $t=\frac{s-i}{(s+i)\gamma^2}$ and $z=1-\frac{2}{\rho^2}$. The resulting expression involves the square root
\begin{align}
    \sqrt{\frac{(s-i)^2}{\gamma ^4 (s+i)^2}-\frac{2 \left(1-\frac{2}{\rho ^2}\right) (s-i)}{\gamma ^2 (s+i)}+1} \nonumber
\end{align}
which simplifies to
\begin{align}
    \left(1-\frac{s-i }{\gamma ^2 (s+i)}\right)\sqrt{1-\frac{s^2+1}{\rho ^2 \left(\frac{1}{2} \left(\gamma +\frac{1}{\gamma }\right)+\frac{1}{2} i \left(\frac{1}{\gamma }-\gamma
   \right) s\right)^2}} \nonumber
\end{align}
this is precisely the $\sqrt{1-\frac{s^2+1}{\xi^2}}$ root which occurs in (\ref{defPhi}) and its derivative, with $\xi=\rho\left(\frac{1}{2} \left(\gamma +\frac{1}{\gamma }\right)+s \frac{i}{2} \left(\frac{1}{\gamma }-\gamma \right)\right)$. From this, (\ref{Fexpand}) and (\ref{Gexpand}) can be obtained through laborious but fairly straightforward algebra.
}

The factors $\rho ^{-m-1} P_{n-1}^{(m,0)}\left(1-\frac{2}{\rho ^2}\right)$ which occur in (\ref{Fexpand}) and (\ref{Gexpand}) have the remarkable property that
\begin{align}
    \int_1^{\infty}\rho \left(\rho ^{-m-1} P_{n-1}^{(m,0)}\left(1-\frac{2}{\rho ^2}\right)\right)^2 d\rho = \frac{1}{2m} \label{eq12}
\end{align}
This can be shown by changing the integration variable to $\chi=1-\frac{2}{\rho^2}$
\begin{align}
    \int_1^{\infty}\rho \left(\rho ^{-m-1} P_{n-1}^{(m,0)}\left(1-\frac{2}{\rho ^2}\right)\right)^2 d\rho \nonumber\\
    = \int_{-1}^{1}2^{-m-1} (1-\chi )^{m-1} P_{n-1}^{(m,0)}(\chi )^2 d\chi \nonumber
\end{align}
Note
\begin{align}
    \frac{d}{d\chi}  (1-\chi )^{m} P_{n-1}^{(m,0)}(\chi )^2 &= -m (1-\chi )^{m-1} P_{n-1}^{(m,0)}(\chi )^2 \nonumber\\ &+ (1-\chi )^{m} 2 P_{n-1}^{(m,0)}(\chi ) \frac{d}{d\chi} P_{n-1}^{(m,0)}(\chi ) \nonumber
\end{align}
After integrating both sides, the second term in the rhs. vanishes due to the Jacobi polynomials' orthogonality under the weight $(1-\chi )^{m}$, and we get
\begin{align}
    -2^{m} P_{n-1}^{(m,0)}(-1 )^2 &= -\int_{-1}^{1}m (1-\chi )^{m-1} P_{n-1}^{(m,0)}(\chi )^2 d\chi 
\end{align}
Now, $P_{n-1}^{(m,0)}(-1 )=(-1)^{n-1}$ so
\begin{align}
    2^{m}  &= \int_{-1}^{1}m (1-\chi )^{m-1} P_{n-1}^{(m,0)}(\chi )^2 d\chi \\
    \frac{1}{2m}  &= \int_{-1}^{1} 2^{-m-1}(1-\chi )^{m-1} P_{n-1}^{(m,0)}(\chi )^2 d\chi  
\end{align}
which proves (\ref{eq12}).

Returning to (\ref{eq6}), the Laurent series which defines the integrand in the annulus is the product of (\ref{Fexpand}),  (\ref{Gexpand}), and $-i/\gamma$. This product converges where all its factors converge, i.e. in the annulus $r_a<|\gamma|<R_a$. To evaluate the countour integral, we only need the $1/\gamma$ term in the Laurent series, which we get from the termwise product of (\ref{Fexpand}) and (\ref{Gexpand}), so (\ref{eq6}) becomes (noting $|\sqrt{-s^2}-1|^2=|s-i|^2$)
\begin{align}
      \oint \frac{-i}{\gamma } \left(\frac{4m^2}{|s-i|^{2m}} \sum_{n=1}^{\infty} \frac{|s-i|^{2(n+m-1)}}{|s+i|^{2n}}  \left( \rho ^{-m-1} P_{n-1}^{(m,0)}\left(1-\frac{2}{\rho ^2}\right)\right)^2 \right) d\gamma \nonumber\\
      = 8\pi m^2 \sum_{n=1}^{\infty} \frac{|s-i|^{2(n-1)}}{|s+i|^{2n}}  \left( \rho ^{-m-1} P_{n-1}^{(m,0)}\left(1-\frac{2}{\rho ^2}\right)\right)^2 \nonumber
\end{align}
Returning to the $\rho$ integral, using (\ref{eq12}),
\begin{align}
    &\int_1^{\infty}\rho \left(8\pi m^2 \sum_{n=0}^{\infty} \frac{|s-i|^{2(n-1)}}{|s+i|^{2n}}  \left( \rho ^{-m-1} P_{n-1}^{(m,0)}\left(1-\frac{2}{\rho ^2}\right)\right)^2 \right) d\rho \nonumber \\
    &= 4\pi m  \sum_{n=1}^{\infty} \frac{|s-i|^{2(n-1)}}{|s+i|^{2n}} \nonumber\\
    &= \frac{4\pi m}{|s-i|^2}  \sum_{n=1}^{\infty} \left|\frac{s-i}{s+i}\right|^{2n} \nonumber\\
    &= \frac{4\pi m}{|s-i|^2}  \frac{\left|\frac{s-i}{s+i}\right|^2}{1-\left|\frac{s-i}{s+i}\right|^2}\nonumber\\
    &= \frac{4\pi m}{|s-i|^2}  \frac{\left|s-i\right|^2}{\left|s+i\right|^2-\left|s-i\right|^2}\nonumber\\
    &=   \frac{4\pi m}{\Re(s)^2+(\Im(s)+1)^2-\Re(s)^2-(\Im(s)-1)^2}\nonumber\\
    &= \frac{\pi m}{\Im(s)}  \nonumber
\end{align}
So the 2D integral outside the source circle is
\begin{align}
    \int_{-\pi}^{\pi} \int_{1}^{\infty} \rho \left|F_{-m} '\left(\rho(\cos(\theta)+s \sin(\theta))\right)\right|^2 d\rho d\theta = \frac{\pi m}{\Im(s)} \label{itgsum}
\end{align}
which is surprisingly independent of $\Re(s)$. We have checked (\ref{itgsum}) against numerical integration in table \ref{table1}.

\begin{table}[]
    \centering
    \begin{tabular}{|c|c|c|c|}
    \hline
        $m$ & $s$ & $\pi m/\Im(s)$ & Numerical \\ \hline
        1 & $1+i$ & $\pi$ & 3.14159\\
        2 & $1+2i$ & $\pi$ & 3.14159\\
        2 & $1+i$ & $2\pi$ & 6.28319\\
        3 & $1+i$ & $3\pi$ & 9.42478\\
        1 & $1+2i$ & $\pi/2$ & 1.5708\\
        3 & $1+4i$ & $3 \pi/4$ & 2.35619\\
        1 & $3+i$ & $\pi$ & 3.14159\\
        1 & $1+i/100$ & $100\pi$ & 314.159\\
        2 & $1+i/100$ & $200\pi$ & 628.319\\ \hline
    \end{tabular}
    \caption{The 2D integral (\ref{itgsum}): exact result $\pi m/\Im(s)$ compared with numerical integration using Mathematica's \texttt{NIntegrate} function. Note the result's independence on $\Re(s)$.}
    \label{table1}
\end{table}

Finally inserting (\ref{itgsum}) back in (\ref{ipol}), the coupled power is
\begin{align}
    \frac{\omega \epsilon_0}{2} |A|^2 (|s|^2 \Im(P)+\Im(S)) \frac{\pi m}{\Im(s)} \label{eq18}
\end{align}
For the collisionless limit, it will be convenient to formulate the denominator in terms of $\Im(s^2)$. Since $\Im(s)=\frac{\Im\left(s^2\right)}{2 \Re(s)}$, the coupled power (\ref{eq18}) equals
\begin{align}
    \frac{\omega \epsilon_0}{2} |A|^2 (|s|^2 \Im(P)+\Im(S)) \frac{2 \pi m \Re(s)}{\Im(s^2)} \label{eq19}
\end{align}
Note that the coupled power (\ref{eq19}) is valid for all collisionalities, from very high collisions all the way down to the collisionless limit. Let us now consider that limit. 
Let $S=S_0+a i$ and $P=P_0+a b i $, with $S_0>0, P_0<0, a >0$ and $b>0$. $S_0$ and $P_0$ are the collisionless (real) values of the Stix parameters. We intend to take the $a\rightarrow 0$ limit, and expect that limit to be independent of $b$ (i.e. independent of the direction along which we approach the limit in ($\Im(S),\Im(P)$) space). Note
\begin{align}
    |s|^2 \Im(P)+\Im(S)=a \left(1-b\frac{S_0}{P_0}\right) + O(a^2)
\end{align}
and
\begin{align}
    s^2 = -\frac{S_0}{P_0}-\frac{i a \left(1-b\frac{S_0}{P_0}\right)}{P_0}+ O(a^2)
\end{align}
So the $b$-dependence in $|s|^2 \Im(P)+\Im(S)$ cancels the $b$-dependence in $\Im(s^2)$: we can indeed take the $\Im(S)\rightarrow 0,\Im(P)\rightarrow 0$ collisionless limit of (\ref{eq19}), and that limit is indeed independent of the direction of approach.
The limit is
\begin{align}
    \lim_{a\rightarrow 0}\frac{\omega \epsilon_0}{2} |A|^2 a \left(1-b\frac{S_0}{P_0}\right) \frac{2 \pi m \Re(s)}{-\frac{a \left(1-b\frac{S_0}{P_0}\right)}{P_0}} = \omega \epsilon_0 |A|^2 \pi m \sqrt{-S_0 P_0}\label{finalEq12}
\end{align}
(\ref{finalEq12}) is the total power dissipated by the resonance cone, in the cold collisionless limit. \textcolor{black}{It is interesting to note that the power (whether collisional or collisionless) scales with the mode number $m$, which suggests that, in order to couple a finite amount of power, the fourier coefficients that define the potential on the surface must decay at least as $m^{-2}$, which confirms that the potential along the surface must be continuous, as expected.}

\section{Conclusion}
We have calculated the power dissipated in the plasma by a 2D cylindrical source emitting a resonance cone, as predicted by the cold plasma theory \textcolor{black}{(using the electrostatic approximation for the resonance cone)}, for any collisionality and any mode number. This power remains finite even in the collisionless limit, even as the field strength itself is singular in that limit. The power in the collisionless limit is independent of the details of the collision mechanism, i.e. independent of the direction along which we approach the limit in ($\Im(S),\Im(P)$) space. This completes the connection between the resonance cone and the lower hybrid resonance \cite{maquet2021analytical}: in both cases, the cold plasma theory predicts singular wave fields and a finite coupled power in the collisionless limit.

In the collisionless limit, the power coupled to the resonance cone scales with $\sqrt{-P S}$. For any collisionality, the power scales with the wave frequency and with the azimuthal mode number $m$. The power is independent of the source radius $r_s$.

We are grateful to an anonymous reviewer of \cite{tierens2024slow}, for drawing our attention to this topic and for predicting the result's independence on $r_s$\textcolor{black}{, by noting that the $r$-dependence in (\ref{ipol}) cancels after switching to the dimensionless radius $\rho=r/r_s$ as integration variable}.

This manuscript has been authored by UT-Battelle, LLC, under Contract No. DE-AC05-00OR22725 with the U.S. Department of Energy (DOE). This material is based upon work supported by the U.S. Department of Energy, Office of Science, Office of Advanced Scientific Computing Research and Office of Fusion Energy Sciences, Scientific Discovery through Advanced Computing (SciDAC) program. The publisher acknowledges the US government license to provide public access under the DOE Public Access Plan (http://energy.gov/downloads/doe-public-access-plan).

%
\bibliography{sample.bib}
%
%
%
%

\end{document}